\documentclass[12pt]{article}
\usepackage{graphicx}
\usepackage{graphics}
\usepackage{fancybox}
\usepackage{amsmath,float,latexsym,psfrag,epsf,epsfig,amssymb,
chicago,rotating,subfigure}
\usepackage{color}

\newcommand{\pref}[1]{%
    \ref{#1} \ifnum\count0=\pageref{#1}\relax%
    \else (page \pageref{#1})\fi}

\newcommand{\eref}[1]{%
        \ref{#1}\ifnum\count0=\pageref{#1}\relax%
        \else {, p.\pageref{#1}}\fi}

\newcommand{\comment}[1]{}

\newenvironment{algorithm}{\vspace{5 mm}\sc}{\vspace{5 mm}}
\newlength{\labwidth}
\newcommand{\step}[1]{%
    \settowidth{\labwidth}{#1\ }%
    \par\noindent%
    \global\hangindent\labwidth {#1}%
    \hbox{ }%
    }%

\newcommand{\bfx}{\hbox{\boldmath$x$}}

\newcommand{\bfy}{\hbox{\boldmath$y$}}

\newcommand{\E}{{\mathbf{E}}}

\begin{document}

\begin{center}

{\Large \bfseries Classification using distance nearest neighbours}
\vspace{5 mm}

{\large N. FRIEL} \\
{\textit{School of Mathematical Sciences, University College Dublin, Ireland.}}\\
and \\{\large A.N. PETTITT} \\
{\textit{Discipline of Mathematical Sciences, Queensland University of Technology, Australia.} }\\

\vspace{5mm}


\end{center}

\bibliographystyle{mybib}

\begin{abstract}

This paper proposes a new probabilistic classification algorithm using a Markov random field
approach. The joint distribution of class labels is explicitly modelled using the distances
between feature vectors. Intuitively, a class label should depend more on class labels which are
closer in the feature space, than those which are further away. Our approach builds on previous
work by Holmes and Adams \citeyear{hol:ada02,hol:ada03} and Cucala \textit{et al}
\citeyear{cuc:mar08}. Our work shares many of the advantages of these approaches in providing
a probabilistic basis for the statistical inference. In comparison to previous work, we present
a more efficient computational algorithm to overcome the intractability of the Markov random
field model. The results of our algorithm are encouraging in comparison to the $k-$nearest
neighbour algorithm.
\end{abstract}

\section{Introduction}

This paper is concerned with the problem of supervised classification, a topic of interest in both statistics and machine learning.
Hastie {\it et al} \citeyear{has:tib:fri00} gives a description of various classification methods. We outline our
problem as follows. We have a collection of training data $\{(x_i,y_i),i=1,\dots,n\}$. The values in the
collection $\bfx = \{x_1,\dots,x_n\}$ are often called features and can be conveniently
thought of as covariates.
We denote the class labels as $\bfy=\{y_1,\dots,y_n\}$, where each $y_i$ takes one of the
values $1,2,\dots,G$. Given a collection of incomplete/unlabelled
test data $\{(x_i,y_i),i=n+1,\dots,n+m\}$, the problem amounts to predicting the class labels
for $\bfy^*=\{y_{n+1},\dots,y_{n+m}\}$ with corresponding feature vectors
$\bfx^*=\{x_{n+1},\dots,x_{n+m}\}$.

Perhaps the most common approach to classification is the well-known $k-$nearest neighbours ($k-$nn) algorithm.
This algorithm amounts to classifying an unlabelled $y_{n+i}$ as the most common class among the
$k$ nearest neighbours of $x_{n+i}$ in the training set $\{(x_i,y_i),i=1,\dots,n\}$.
While this algorithm is easy to implement, and often gives good performance, it can be
criticised since it does not allow any uncertainty to be associated to the test class labels, and
to the value to $k$. Indeed the choice of $k$ is crucial to the performance of the algorithm. The value of
$k$ is often chosen on the basis of leave-one-out cross-validation.


There has been some interest in extending the $k-$nearest neighbours algorithm to allow for
uncertainty in the test class labelling, most notably by \cite{hol:ada02}, \cite{hol:ada03} and
more recently \shortcite{cuc:mar08}. Each of these probabilistic variants of the $k-$nearest neighbour
algorithm, is based on defining a neighbourhood of each point $x_i$, consisting of the $k$ nearest neighbours
of $x_i$. But moreover, each of these neighbouring points has equal influence in determining the missing
class label for $y_i$, regardless of distance from $x_i$. In this article we present a class of models,
the \textit{distance nearest neighbour} model, which shares many of the advantages of these probabilistic approaches, but in contrast to these approaches, the relative influence of neighbouring points depends
on the distance from $x_i$. Formally, the distance nearest neighbour model is a discrete-valued Markov random field, and, as is typical with such models, depends on an intractable normalising constant. To overcome this problem we use the exchange algorithm of Murray {\textit{et al.}} \citeyear{Murray06} and illustrate that this provides a computationally efficient algorithm with very good mixing properties.
This contrasts with the difficulties encountered by Cucala \textit{et al.} \citeyear{cuc:mar08} in their implementation of the sampling scheme of M{\o}ller \textit{et al} \citeyear{mol:pet06}.

This article is organised as follows. Section 2 presents a recent overview of recent probabilistic approaches
to supervised classification. Section 3 introduces the new distance nearest neighbour model and outlines
how it compares and contrasts to previous probabilistic nearest neighbour approaches. We provide a computationally
efficient framework for carrying out inference for the distance nearest neighbour model in Section 4. The performance
of the algorithm is illustrated in Section 5 for a variety of benchmark datasets, as well as challenging
high-dimensional datasets. Finally, we present some closing remarks in Section 6.

\section{Probabilistic nearest neighbour models}

Holmes and Adams \citeyear{hol:ada03} attempted to place the $k-$nn algorithm in a probabilistic setting
therefore allowing for uncertainty in the test class labelling. In their approach the full-conditional
distribution for a training label is written as
\[ \pi(y_i|\bfy_{-i},\bfx,\beta,k) \propto \exp\left( \beta\sum_{j\sim^k i} I(y_i=y_j)/k \right), \]
where the summation is over the $k$ nearest neighbours of $x_i$ and where $I(y_i=y_j)$ is an indicator
function taking the value $1$ if $y_i=y_j$ and $0$ otherwise. The notation, $j\sim^k i$ means that
$x_j$ is one of the $k$ nearest neighbours of $x_i$.
However, as pointed out in \shortcite{cuc:mar08}, there is a difficulty with this formulation, namely that
there will almost never be a joint probability for $\bfy$ corresponding to this collection of
full-conditionals. The reason is simply because the $k-$nn neighbourhood system is usually asymmetric.
If $x_i$ is one of the $k$ nearest neighbours of $x_j$, then it does not necessarily follow that
$x_j$ is one of the $k$ nearest neighbours of $x_i$.

Cucala \textit{et al.} \citeyear{cuc:mar08} corrected the issue surrounding the asymmetry of the $k$-nn
neighbourhood system. In their probabilistic $k$-nn ($pk$-nn) model, the full-conditional for class label $y_i$ appears as
\begin{equation}
 \pi(y_i|\bfy_{-i},\bfx,\beta,k) \propto \exp\left( \beta/k
    \left\{\sum_{j\sim^k i} I(y_i=y_j) + \sum_{i\sim^k j} I(y_i=y_j) \right\} \right),
\label{eqn:cuc}
\end{equation}
and this gives rise to the joint distribution
\[ \pi(\bfy|\bfx,\beta,k) \propto \exp\left( \beta/k \sum_{i=1}^n \sum_{j\sim^k i} I(y_i=y_j) \right). \]
Therefore under this model, following~(\ref{eqn:cuc}), mutual neighbours are given double weight, with respect to non-mutual neighbours and for this reason the model could be seen, perhaps, as an ad-hoc solution to this problem.

It is important to also note that both Holmes and Adams \citeyear{hol:ada02} and Cucala \textit{et al.} \citeyear{cuc:mar08}
allow the value of $k$ to be a variable. Therefore the neighbourhood size can vary. Holmes and Adams \citeyear{hol:ada02}
argue that allowing $k$ to vary has a certain type of smoothing effect.

\section{Distance nearest neighbours}
\label{sec:dnn}

Motivated by the work of Holmes and Adams \citeyear{hol:ada02} and Cucala \textit{et al.} \citeyear{cuc:mar08} our
interest focuses on modelling the distribution of the training data as a Markov random field. Similar to these
approaches, we consider a Markov random field based approach, but in contrast our approach
explicitly models and depends on the distances between points in the training set. Specifically,
we define the full-conditional distribution of the class label $y_i$ as
\[ \pi(y_i|\bfy_{-i},\bfx,\beta,\sigma) \propto \exp\left( \beta\sum_{j=1, j\neq i}^n
  w_j^iI(y_j=y_i)  \right). \]
Positive values of the Markov random field parameter $\beta$ encourage aggregation of the
class label. When $\beta=0$, the class labels are uncorrelated.
In contrast to the $pk$-nn model, here the neighbourhood set of $x_i$ is constructed to be
\[ \bfx\setminus\{x_i\} = \{x_1,\dots,x_{i-1},x_{i+1},\dots,x_n\} \]
and is therefore of maximal size. We consider three possible models depending on how
the collection of weights $\{w_i^j\}$ for $j=1,\dots,i-1,i+1,\dots,n$ are defined.

\begin{enumerate}
\item $d-$nn$_1$:
\[ w_i^j \propto \exp\left\{ -\frac{d(x_i,x_j)^2}{2\sigma^2}\right\},\; \mbox{for}\; j=1,\dots,i-1,i+1,\dots,n,  \]
where $d$ is a distance measure such as Euclidean. 
\item $d-$nn$_2$:
\[ w_i^j \propto \epsilon + (1-\epsilon) I(d(x_i,x_j)<\sigma),\; \mbox{for}\; j=1,\dots,i-1,i+1,\dots,n,  \]
again where, $I$ is an indicator function taking value $1$, if $d(x_i,x_j)<\sigma$ and $0$, otherwise.
Further, $\epsilon\in (0,1)$ is defined as a constant, and is set to a value close to $0$. (Throughout
this paper we assign the value $\epsilon=10^{-10}$.)
A non-zero value of $\epsilon$ guarantees that if there are no features within a distance $\sigma$ of $x_i$ then the class of $y_i$ is modelled using the marginal proportions of the class labels.
\item $d-$nn$_3$:
\[ w_i^j \propto \exp\left\{ -d(x_i,x_j)\sigma \right\},\; \mbox{for}\; j=1,\dots,i-1,i+1,\dots,n.  \]
\end{enumerate}

Clearly the neighbour system for both models is symmetric, and so the Hammersley-Clifford theorem guarantees that the joint distribution of the class
labels is a Markov random field. This joint distribution is written as
\begin{equation}
 \pi(\bfy|\bfx,\beta,\sigma) = \frac{q(\bfy|\beta,\sigma,\bfx)}{z(\beta,\sigma)} =
 \frac{\exp\left( \beta\sum_i \sum_{j=1, j\neq i}^n w_j^iI(y_j=y_i) \right) }{z(\beta,\sigma)}.
 \label{eqn:joint}
\end{equation}

As usual, the normalising constant of such a Markov random field is difficult to evaluate in all but
trivial cases. It appears as
\begin{equation}
 z(\beta,\sigma) = \sum_{y_1}\dots\sum_{y_n} \exp\left( \beta\sum_i
  \sum_{j=1, j\neq i}^n w_j^iI(y_j=y_i)  \right).
\label{eqn:nc}
\end{equation}
Some comments:
\begin{enumerate}
 \item The $k-$nn algorithm and its probabilistic variants always contain neighbourhoods of size $k$, regardless
of how far each of neighbouring points are from the center point, $x_i$. Moreover, each neighbouring point $x_j$
has equal influence, regardless of distance from $x_i$. It could therefore be argued that these algorithms are
not sensitive to outliers. By contrast the distance nearest neighbour models deal with outlying points in a more
robust manner, since if a point $x_j$ lies further away from other neighbours of $x_i$, then it will have a relatively
smaller weight, and consequently less influence in determining the likely class label of $y_i$.

\item The formulation of distance nearest neighbour models includes every training point in the neighbourhood set, but the value of $\sigma$
determines the relative influence of points in the neighbourhood set. For the $d-$nn$_1$ model, small values of $\sigma$ imply that only those points with small distance from the centre point will be influential, while for large values of $\sigma$, points in the neighbourhood set are more uniformly weighted. Similarly, for the $d-$nn$_2$ model, points within a $\sigma$ radius of the center point
are weighted equally, while those outside a $\sigma$ radius of the center point will have relatively
little weight, when $\epsilon$ is very close to $0$.
By contrast, for the $d-$nn$_3$ model, large values of the parameter
$\sigma$ imply that points close to the centre point will be influential.
\item For the $d-$nn$_2$ model, if there are no features in the training set within a distance $\sigma$
of $x_i$, then
\[ \pi(y_i=j|\bfy_{-i},\bfx,\beta,\sigma) \propto \exp\left( \beta p_j^i \right),\; \mbox{for}\; j=1,\dots,G,  \]
where $p_j^i$ denotes the proportion of class labels $j$ in the set $\bfy\setminus\{y_i\}$.
The parameter $\beta$ determines the dependence on the class proportions.  A large value of $\beta$ typically predicts the class label to be the class with the largest proportion, whereas a small value of $\beta$ results in a prediction which is almost uniform over all possible classes.
Conversely,
if there any feature vectors within a radius $\sigma$ of $x_i$, then the class labels for these features
will most influence the class label of $y_i$.
\item As $\beta\rightarrow\infty$, the most frequently occurring training label in the neighbourhood
of a test point will be chosen with increasing large probability. The $\beta$ parameter can be thought of, in a
sense, as a tempering parameter. In the limit as $\beta\rightarrow\infty$, the modal class label in the neighbourhood
set has probability $1$.

\end{enumerate}

There has been work on extending the $k-$nearest neighbours algorithm to weight neighbours within the neighbourhood of size $k$.  For example, \cite{dud75} weighted neighbours using the distance in a linear manner while standardizing weights to lie in $[0,1]$.

A model similar to the $d-$nn$_1$ model appeared in \cite{zhu:gha02}, but it does not contain
the $\beta$ Markov random field parameter to control the level of aggregation in the spatial
field. Moreover, the authors outline some MCMC approaches, but note that inference for this model is
challenging. The aim of this paper is to illustrate how this model may be generalised and to
illustrate an efficient algorithm to sample from this model. We now address the latter issue.

\section{Implementing the distance-nearest neighbours algorithm}

Throughout we consider a Bayesian treatment of this problem. The posterior distribution of test
labels and Markov random field parameters can be expressed as
\begin{displaymath}
 \pi(\bfy^*,\beta,\sigma|\bfy,\bfx,\bfx^*) \propto \pi(\bfy,\bfy^*|\beta,\sigma,\bfx,\bfx^*)\pi(\beta)\pi(\sigma),
\end{displaymath}
where $\pi(\beta)$ and $\pi(\sigma)$ are prior distributions for $\beta$ and $\sigma$, respectively.
Note, however that the first term on the right hand side above depends on the intractable normalising
constant (\ref{eqn:nc}). In fact, the number of test labels is often much greater than the number of
training labels, and so the resulting normalising constant for the distribution $\pi(\bfy,\bfy^*|\beta,\sigma,\bfx,\bfx^*)$
involves a summation over $G^{n+m}$ terms, where as before $n,m$ and $G$ are the number of test data points, training
data points and class labels, respectively. A more pragmatic alternative is to consider the posterior distribution of the unknown parameters for
the training class labels,
\begin{displaymath}
 \pi( \beta,\sigma|\bfx,\bfy) \propto \pi(\bfy|\beta,\sigma,\bfx) \pi(\beta) \pi(\sigma),
\end{displaymath}
where now the normalising constant depends on $G^n$ terms. Test class labels can then be predicted by averaging
over the posterior distribution of the training data,
\begin{displaymath}
 \pi( y_{n+i}|x_{n+i},\bfx,\bfy) = \int \pi(y_{n+i}|x_{n+i},\bfx,\bfy,\beta,\sigma) \pi( \beta,\sigma|\bfx,\bfy) d\beta d\sigma.
\end{displaymath}
Obviously, this assumes that the test class labels, $\bfy^*$ are mutually independent, given the training data, which
will typically be an unreasonable assumption.
The training class labels are modelled as being mutually independent. Clearly, this is not ideal
from the Bayesian perspective.
Nevertheless, it should reduce the computational complexity of the
problem dramatically.

In practice, we can estimate the predictive probability of $y_{n+i}$ as an ergodic average
\begin{displaymath}
 \pi( y_{n+i}|x_{n+i},\bfx,\bfy) \approx \frac{1}{J}\sum_{j=1}^J \pi(y_{n+i}|x_{n+i},\bfx,\bfy,\beta^{(j)},\sigma^{(j)}),
\end{displaymath}
where $\beta^{(j)},\sigma^{(j)}$ are samples from the posterior distribution $\pi( \beta,\sigma|\bfx,\bfy)$.

\subsection{Pseudolikelihood estimation}

A standard approach to approximate the distribution of a Markov random field is to use a pseudolikelihood
approximation, first proposed in \cite{bes74}. This approximation consists of a product of easily normalised
full-conditional distributions. For our model, we can write a pseudolikelihood approximation as
\begin{displaymath}
 \pi(\bfy|\bfx,\beta,\sigma) \approx \prod_{i=1}^n \pi(y_i|\bfy_{-i},\bfx,\beta,\sigma) =
\prod_{i=1}^n \frac{\exp\left( \beta\sum_{j=1, j\neq i}^n w_j^iI(y_j=y_i) \right) }
{\sum_{k=1}^G \exp\left( \beta\sum_{j=1, j\neq i}^n w_j^iI(y_j=k) \right) }.
\end{displaymath}
This approximation yields a fast approximation to the posterior distribution, however it does ignore dependencies
beyond first order.

\subsection{The exchange algorithm}

The main computational burden is sampling from the posterior distribution
\begin{eqnarray*}
 \pi( \beta,\sigma | \bfx,\bfy ) &\propto& \pi(\bfy|\beta,\sigma,\bfx) \pi(\beta) \pi(\sigma) \\
 &=& \frac{q(\bfy|\beta,\sigma,\bfx)}{z(\beta,\sigma)} \pi(\beta) \pi(\sigma).
\end{eqnarray*}
A naive implementation of a Metropolis-Hastings algorithm proposing to move from $(\beta,\sigma)$ to
$(\beta',\sigma')$ would require calculation of the following ratio at each sweep of the algorithm
\begin{equation}
 \frac{q(\bfy|\beta',\sigma',\bfx)\pi(\beta')\pi(\sigma')}{q(\bfy|\beta,\sigma,\bfx)\pi(\beta)\pi(\sigma)}
 \times \frac{z(\beta,\sigma)}{z(\beta',\sigma')}.
\label{eqn:MHratio}
\end{equation}
The intractability of the normalising constants, $z(\beta,\sigma)$ and $z(\beta',\sigma')$, makes this
algorithm unworkable. There has been work which has tackled the problem of sampling from such complicated
distributions, for example, \shortcite{mol:pet06}. The algorithm presented in this paper overcomes the
problem of sampling from a distribution with intractable normalising constant, to a large extent. However
the algorithm can result in an MCMC chain with poor mixing among the parameters. The algorithm in
\shortcite{mol:pet06} has been extended and improved in \cite{Murray06}.

The algorithm samples from an augmented distribution
\begin{equation}
 \pi(\beta',\sigma',\bfy',\sigma,\beta|\bfy,\bfx) \propto \pi(\bfy|\beta,\sigma,\bfx)\pi(\beta)\pi(\sigma)
 h(\beta',\sigma'|\beta,\sigma) \pi(\bfy'|\beta',\sigma',\bfx),
\label{eqn:exchange}
\end{equation}
where $\pi(\bfy'|\beta',\sigma',\bfx)$ is the same distance nearest-neighbour distribution as the
training data $\bfy$. The distribution $h(\beta',\sigma'|\beta,\sigma)$ is any arbitrary distribution
for the augmented variables $(\beta',\sigma')$ which might depend on the variables $(\beta,\sigma)$, for
example, a random walk distribution centred at $(\beta,\sigma)$. It is clear that the marginal distribution
of (\ref{eqn:exchange}) for variables $\sigma$ and $\beta$ is the posterior distribution of interest.

The algorithm can be written in the following concise way:
\begin{algorithm}
\step{1.} Gibbs update of $(\beta',\sigma',\bfy')$:\\
(i) Draw $(\beta',\sigma') \sim h(\cdot,\cdot|\beta,\sigma)$.\\
(ii) Draw $\bfy' \sim \pi(\cdot|\beta',\sigma',\bfx)$.
\step{2.} Propose to move from $(\beta,\sigma,\bfy),(\beta',\sigma',\bfy')$ to $(\beta',\sigma',\bfy),(\beta,\sigma,\bfy')$.
(Exchange move) with probability
\begin{displaymath}
 \min\left( 1, \frac{q(\bfy'|\beta,\sigma,\bfx)\pi(\beta')\pi(\sigma')h(\beta,\sigma|\beta',\sigma')q(\bfy|\beta',\sigma',\bfx)}
{q(\bfy|\beta,\sigma,\bfx)\pi(\beta)\pi(\sigma)h(\beta',\sigma'|\beta,\sigma)q(\bfy'|\beta',\sigma',\bfx)}
\times \frac{z(\beta,\sigma)z(\beta',\sigma')}{z(\beta,\sigma)z(\beta',\sigma')} \right).
\end{displaymath}
\end{algorithm}

Notice in Step 2, that all intractable normalising constants cancel above and below the fraction.
The difficult step of the algorithm in the context of the $d-$nn model is Step 1 (ii), since this requires a draw from $\pi(\bfy'|\beta',\sigma',\bfx)$.
Perfect sampling \cite{pro:wil96} is often possible for Markov random field models, however a pragmatic alternative
is to sample from $\pi(\cdot|\beta',\sigma',\bfx)$ by standard MCMC methods, for example, Gibbs sampling, and take
a realisation from a long run of the chain as an approximate draw from the distribution.
Note that this is the approach that Cucala \textit{et al.} \citeyear{cuc:mar08}
take. They argue that perfect sampling is possible for the $pk-$nn algorithm
for the case where there are two classes, but that the time to coalescence can be prohibitively large. They note that
perfect sampling for more than two classes is not yet available.

Note that this algorithm has some similarities with Approximate Bayesian Computation (ABC)
methods \cite{sis:fan:tan07} in the sense that ABC algorithms also rely on drawing exact values from 
analytically intractable distributions. By contrast however, ABC algorithms rely on comparing summary statistics of the auxiliary
data to summary statistics of the observed data.
Finally, note that the Metropolis-Hastings ratio in step $2$ above, after re-arranging some terms, and assuming that
$h(\beta,\sigma|\beta',\sigma')$ is symmetric can be written as
\[
 \frac{q(\bfy|\beta',\sigma',\bfx)\pi(\beta')\pi(\sigma')q(\bfy'|\beta,\sigma,\bfx)}
{q(\bfy|\beta,\sigma,\bfx)\pi(\beta)\pi(\sigma)q(\bfy'|\beta',\sigma',\bfx)}.
\]
Comparing this to~(\ref{eqn:MHratio}), we see that the ratio of normalising constants, $z(\beta,\sigma)/z(\beta',\sigma')$,
is replaced by $q(\bfy'|\beta,\sigma,\bfx)/q(\bfy'|\beta',\sigma',\bfx)$, which itself can be interpreted as an importance
sampling estimate of $z(\beta,\sigma)/z(\beta',\sigma')$, since
\[ \E_{\bfy'|\beta',\sigma'} \Big[\frac{q(\bfy'|\beta,\sigma,\bfx)}{q(\bfy'|\beta',\sigma',\bfx)}\Big] =
 \int \frac{q(\bfy'|\beta,\sigma,\bfx)}{q(\bfy'|\beta',\sigma',\bfx)} \frac{q(\bfy'|\beta',\sigma',\bfx)}{z(\beta',\sigma')}\;d\bfy'
 = \frac{z(\beta,\sigma)}{z(\beta',\sigma')}.
\]

\section{Results}

The performance of our algorithm is illustrated in a variety of settings. We begin by testing
the algorithm on a collection of benchmark datasets and follow this by exploring two real
datasets with high-dimensional feature vectors. Matlab computer code and all of the datasets
(test and training) used in this paper can be found at \texttt{mathsci.ucd.ie/$\sim$nial/dnn/}.

\subsection{Benchmark datasets}

In this section we present results for our model and in each case we compare results with the $k-$nn algorithm
for well known benchmark datasets.
A summary description of each dataset is presented in Table~\ref{tab:bench}.

\begin{table}[htp]
\begin{center}
\begin{tabular}{l|ccc}
       & $G$ & $F$ & $N$ \\
\hline
Pima   & $2$  & $8$ & $532$ \\
Forensic glass & $4$  & $9$ & $214$ \\
Iris   & $3$   & $4$  & $150$  \\
Crabs  & $4$  & $5$ & $200$ \\
Wine   & $3$   & $13$  & $178$ \\
Olive  & $3$   & $9$  & $572$ \\
\hline
\end{tabular}
\end{center}
\caption{Summary of the benchmark datasets: $G,F,N$ correspond to the number of classes, the number
of features and the overall number of observations, respectively.}
\label{tab:bench}
\end{table}

In all situations, the training dataset was approximately $25\%$ of the size of the overall dataset,
thereby presenting a challenging scenario for the various algorithms. Note that the sizes of each dataset
ranges from quite small in the case of the iris dataset, to reasonably large in the case of the forensic
dataset. In all examples, the data was standardised to give transformed features with zero mean and unit variance.
In the Bayesian model, non-informative $N(0,50^2)$ and $U(0,100)$ priors were chosen for $\beta$ and $\sigma$,
respectively. Each $d-$nn algorithm was run for $20,000$ iterations, with the first $10,000$ serving as
burn-in iterations. The auxiliary chain within the exchange algorithm was run for $1,000$ iterations.
The $k-$nn algorithm was computed for values of $k$ from $1$ to half the number of features in the training
set. In terms of computational run time, the $d-$nn algorithms took, depending on the size of the dataset, between
$1$ to $12$ hours to run using Matlab code on a $2$GHz desktop machine.

A summary of misclassification error rates is presented in Table~\ref{tab:bench_res} for various benchmark datasets. In almost all of the situations $d-$nn$_1$ and $d-$nn$_3$ performs at least as well as $k-$nn and often considerably better. In general, $d-$nn$_1$ and $d-$nn$_3$ performed better than $d-$nn$_2$.
A possible explanation for this may be due to the
cut-off nature of the weight function in the $d-$nn$_2$ model, since if a point $x_i$ has no neighbours inside a ball of
radius $\sigma$, then $w_i^j$ is uniform over the entire test set, and consequently there is no effect of distance. By contrast,
both the $d-$nn$_1$ and $d-$nn$_3$ models, have weight functions which depend on distance, and smoothly converge to a uniform
distribution as $\sigma\rightarrow \infty$ and $\sigma\rightarrow 0$, respectively.

\begin{table}[htp]
\begin{center}
\begin{tabular}{l|cccc}
  & $k-$nn & $d-$nn$_1$ & $d-$nn$_{2}$ & $d-$nn$_{3}$ \\
\hline
Pima   & $30\%$  & $29\%$ & $32\%$ & $30\%$\\
Forensic glass & $35\%$  & $33\%$ & $39\%$ & $31\%$ \\
Iris   & $6\%$   & $5\%$  & $5\%$ & $6\%$ \\
Crabs  & $16\%$  & $16\%$ & $23\%$ & $16\%$ \\
Wine   & $6\%$   & $4\%$  & $6\%$ &  $4\%$    \\
Olive  & $1\%$   & $3\%$  &$4\%$ & $2\%$\\
\hline
\end{tabular}
\end{center}
\caption{Misclassification error rates for various benchmark dataset. The value of $k$ in the $k-$nn algorithm
was chosen as the value that minimises the leave-one-out cross-validation error rate. (In the case of a tie,
the smallest value of $k$ was selected.)}
\label{tab:bench_res}
\end{table}

\subsection{Classification with large feature sets: food authenticity}

Here we consider two datasets concerned with food authentication. The first example involves
samples of Greek olive oil from $3$ different regions, and the second example involves
samples of $5$ different types of meat. In both situations each sample was analysed using
near infra-red spectroscopy giving rise to $1050$ reflectance values for wavelengths in the
range $400-2098$nm. These $1050$ reflectance values serve as the feature vector for each sample.
The objective in both examples is to authenticate a test sample based on a training set of
complete data (both reflectance values and class labels). Details of how both datasets were collected
appear in \shortcite{mcel:dow:fea99}, and were analysed using a model-based clustering approach in \shortcite{dea:mur:dow06}.

\subsubsection{Classifying meat samples}

Here $231$ samples of meat were collected.
The aim of this study was to see if these measurements could be used to classify each meat sample according to whether it is chicken, turkey, pork, beef or lamb. The data were randomly split into $60$ training samples and $171$ test samples. The respective number of samples in each class is given in the table below.
\begin{table}[htp]
\begin{center}
\begin{tabular}{ccc}
& Training & Test \\
\hline
Chicken & 15 & 40 \\
Turkey & 20 & 35 \\
Pork & 13 & 42 \\
Beef & 11 & 21 \\
Lamb & 11 & 23 \\
\end{tabular}
\end{center}
\caption{Number of samples within each class for both the training and test datasets}
\end{table}

As before, non-informative normal, $N(0,50^2)$ and uniform $U(0,10)$ priors were chosen for $\beta$ and $\sigma$,
respectively. In the exchange
algorithm, the auxiliary chain was run for $1000$ iterations, and the overall chain ran for $20,000$ of which
the first $10,000$ were discarded as burn-in iterations. The overall acceptance rate for the exchange algorithm was around $25\%$ for each of the $d-$nn models.

\begin{figure}[htp]
\begin{center}
 \hspace*{-3cm}\includegraphics[scale=0.95]{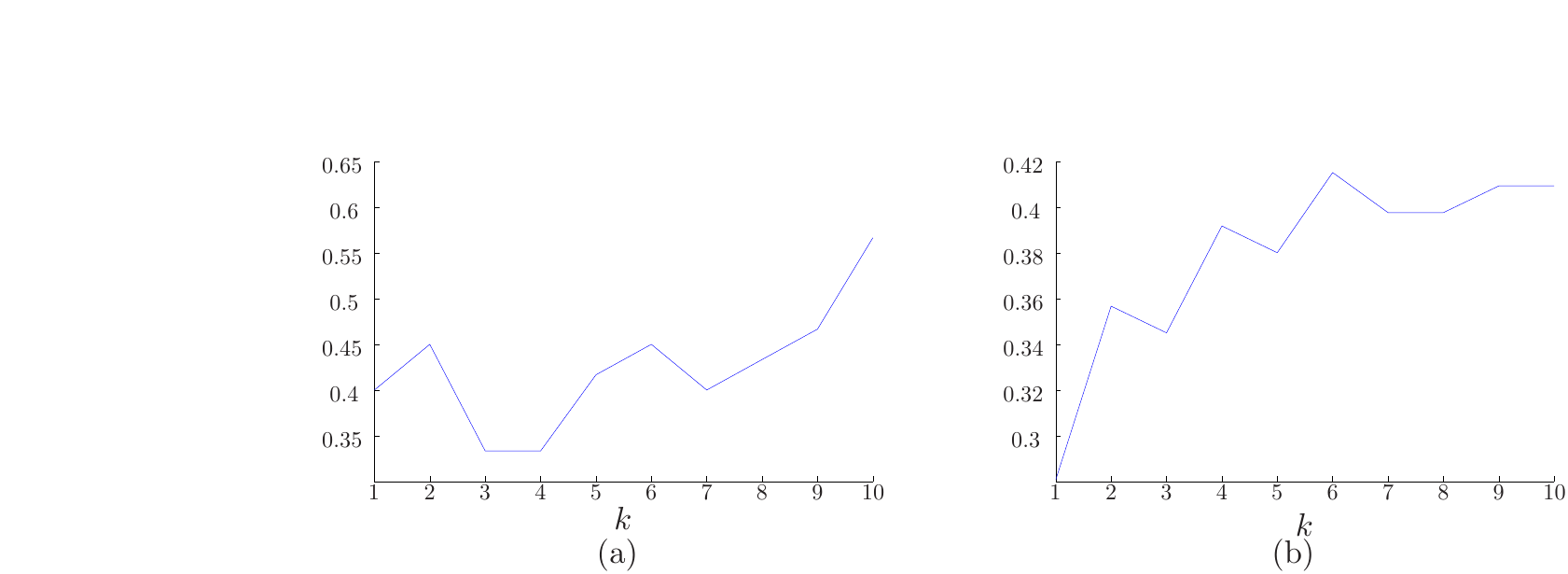}
 \caption{Meat dataset: (a) Training data: misclassification rates of leave-one-out cross-validation for $k-$nn algorithm for varying values of $k$. (b) Test data: misclassification rates for $k-$nn algorithm for varying values of $k$.}
\label{fig:meat_crossval}
\end{center}
\end{figure}

The misclassification error rate for leave-one-out cross-validation on the training dataset is minimised for
$k=3$ and $k=4$. See Figure~\ref{fig:meat_crossval} (a). At both of these values, the $k-$nn algorithm yielded a misclassification error rate of $35\%$ and $39\%$, respectively, for the test dataset.
See Figure~\ref{fig:meat_crossval} (b). By comparison, the $d-$nn$_1$, $d-$nn$_2$  and $d-$nn$_3$ models
achieved misclassification error rates of $29\%$, $33\%$ and $27\%$,
respectively, for the test dataset.
This example further illustrates the value of the $d-$nn models.

%
%
%
%
%
%

\subsubsection{Classifying Greek olive oil}

This example concerns classifying Greek oil samples, again based on infra-red spectroscopy. Here
$65$ samples of Greek virgin olive-oil were collected.
The aim of this study was to see if these measurements could be used to classify each olive-oil sample to
the correct geographical region. Here there were $3$ possible classes (Crete (18 locations), Peloponnese
(28 locations) and other regions (19 locations).

In our experiment the data were randomly split into a training set of $25$ observations and a test set of $40$ observations. In the training dataset the proportion of class labels was similar to that in the complete dataset.

In the Bayesian model,  non-informative $N(0,50^2)$ and $U(0,100)$ priors were chosen for
$\beta$ and $\sigma$. In the exchange algorithm, the auxiliary chain was run for $1000$ iterations, and the overall chain ran for $50,000$ of which the first $20,000$ were discarded as burn-in iterations. The overall acceptance rate
for the exchange algorithm was around $15\%$ for each of the Markov chains.

\begin{figure}[htp]
\begin{center}
 \hspace*{-3cm}\includegraphics[scale=0.95]{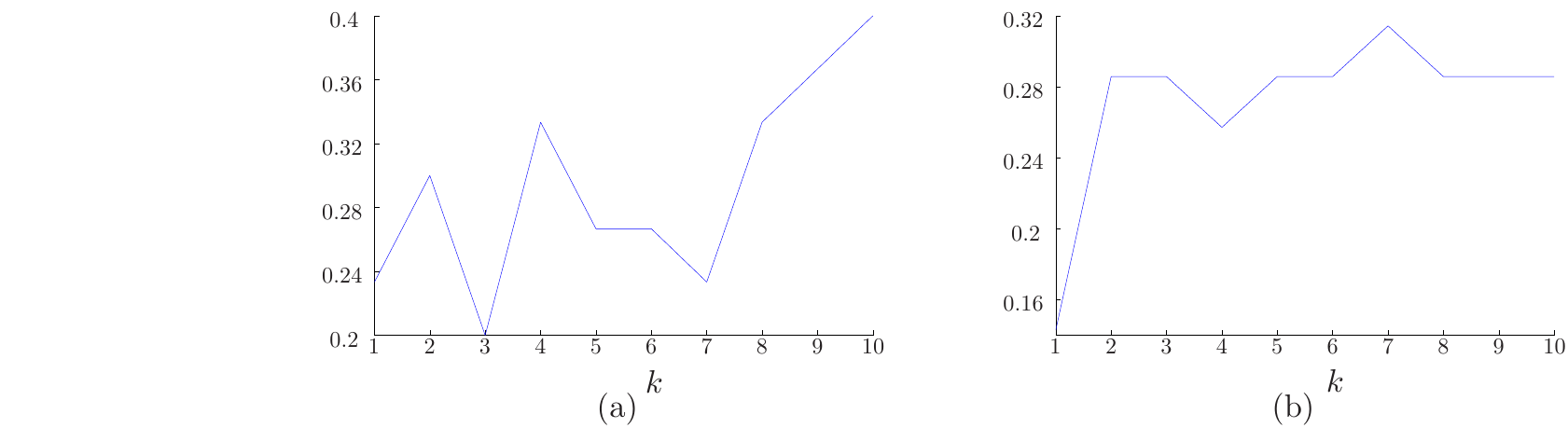}
 \caption{Olive oil dataset: (a) Training data: misclassification rates of leave-one-out cross-validation for $k-$nn algorithm for varying values of $k$. (b) Test data: misclassification rates for $k-$nn algorithm for varying values of $k$.}
  \label{fig:olive_crossval}
\end{center}
\end{figure}

The $d-$nn$_1$, $d-$nn$_2$ and $d-$nn$_3$ models achieved misclassification rates
of $20\%$, $26\%$ and $20\%$, respectively.
In terms of comparison with the $k-$nn algorithm, leave-one-out cross-validation was minimised for
$k=3$ for the training dataset. See Figure~\ref{fig:olive_crossval} (a). The misclassification rates at
this value of $k$ was $29\%$ for the test dataset. See Figure~\ref{fig:olive_crossval} (b).

%
%
%
%
%

It is again encouraging that the $d-$nn algorithms yielded  improved misclassification rates by
comparison.

\section{Concluding remarks}
\label{sec:conc}

In terms of providing a probabilistic approach  to a Bayesian analysis of supervised learning, our work builds on that of Cucala \textit{et al} \citeyear{cuc:mar08} and shares many of the advantages of the approach there, providing a sound setting for Bayesian inference.  The most likely allocations for the test dataset can be evaluated and also the uncertainty that goes with them. So this makes it possible to determine regions where allocation to specific classes is uncertain. In addition, the Bayesian framework allows for an automatic approach to choosing weights for neighbours or neighbourhood sizes.

The present paper also addresses the computational difficulties related to the well-known issue of the intractable normalising constant for discrete exponential family models. While Cucala \textit{et al} \citeyear{cuc:mar08}
demonstrated that MCMC sampling is a practical  alternative to the perfect sampling scheme of M{\o}ller \textit{et al} \citeyear{mol:pet06}, there remain difficulties with their implementation of the approach of \shortcite{mol:pet06}, namely the choice of an auxiliary distribution.  To partially overcome the difficulties of a poor choice, Cucala \textit{et al} \citeyear{cuc:mar08} use an adaptive algorithm where the auxiliary distribution is defined by using historical values in the Monte Carlo algorithm. We use an alternative approach based on the exchange algorithm which avoids this choice or adaptation and has very good mixing properties and therefore also has
computational efficiency.

An issue with the neighbourhood model of Cucala \textit{et al} \citeyear{cuc:mar08}, which is an Ising or Boltzmann type model, is that it is necessary to define an upper value for the association parameter $\beta$.  This parameter value arises from the phase change of the model and which is known for a regular neighbourhood structure but has to be investigated empirically for the probabilistic neighbourhood model.  Our distance nearest neighbour models avoid this difficulty.

Our approach is robust to outliers whereas the nearest neighbour approaches will always have an outlying point having neighbours and therefore classified according to assumed independent distant points which are the nearest neighbours.

\paragraph*{Acknowledgements:} Nial Friel was supported by a Science Foundation Ireland Research Frontiers
Program grant, 09/RFP/MTH2199. Tony Pettitt's research was supported by an Australian Research Council Discovery Grant.

\bibliography{hmrf}

\end{document}